\documentclass[a4paper, amsfonts, amssymb, amsmath, reprint, showkeys, nofootinbib, twoside]{revtex4-1}

\usepackage[english]{babel}
\usepackage[utf8]{inputenc}
\usepackage[colorinlistoftodos, color=green!40, prependcaption]{todonotes}
\usepackage{amsthm}
\usepackage{mathtools}
\usepackage{physics}
\usepackage{xcolor}
\usepackage{graphicx}
\usepackage[left=23mm,right=13mm,top=35mm,columnsep=15pt]{geometry} 
\usepackage{adjustbox}
\usepackage{placeins}
\usepackage[T1]{fontenc}
\usepackage{lipsum}
\usepackage{csquotes}

\usepackage{subfigure}
\usepackage[pdftex, pdftitle={Article}, pdfauthor={Author}]{hyperref} 
\begin{document}
\title{Primordial Magnetic Fields in Light of Dark Ages Global 21-cm Signal}

\author{Vivekanand Mohapatra}
    \email[Email address: ]{p22ph003@nitm.ac.in}
    \affiliation{Department of Physics, National Institute of Technology Meghalaya, Shillong, Meghalaya, India}

\author{Pravin Kumar Natwariya}
    \email[Email address: ]{pvn.sps@gmail.com}
    \affiliation{School of Fundamental Physics and Mathematical Sciences, Hangzhou Institute for Advanced Study, UCAS, Hangzhou 310024, China}
    \affiliation{University of Chinese Academy of Sciences, Beijing 100049, China}

\author{Alekha C. Nayak}
    \email[Email address: ]{alekhanayak@nitm.ac.in}
    \affiliation{Department of Physics, National Institute of Technology Meghalaya, Shillong, Meghalaya, India}

\date{\today} 

\begin{abstract}
We study the constraints on primordial magnetic fields (PMFs) in light of the global 21-cm signal observed during the dark ages. Primordial magnetic fields can heat the intergalactic medium (IGM) via magnetohydrodynamic effects. We investigate the impact of magnetic heating on the Dark Ages global 21-cm signal and constrain the present-day strength of primordial magnetic fields and their spectral indices. Since there were no stars during the Dark Ages, measuring the global 21-cm signal can provide pristine cosmological information. However, detecting this signal using ground-based telescopes is challenging. Several lunar and space-based experiments, such as FARSIDE, DAPPER, and FarView, have been proposed to detect the signal in future. Our findings indicate that measuring the 21 cm global signal during the Dark Ages can provide stronger bounds compared to the existing constraints from Planck 2016. Specifically, the bounds are independent of astrophysical uncertainties and stronger for spectral indices $-2.84\leq n_B\leq -1.58$. Additionally, we explore the dark-ages consistency ratio, which can identify any non-standard heating of the IGM by measuring the 21-cm signal at only three different redshifts. This approach could complement future experiments aimed at detecting the Dark Ages global 21-cm signal.
\end{abstract}


\maketitle
\section{Introduction}\label{sec1}

After the end of the recombination epoch around the redshift of $z\approx 1100$, the primordial plasma predominantly contained neutral hydrogen gas, residual free electrons, protons, and cosmic microwave background photons (CMB). Due to the hyperfine interaction, the ground state of the neutral hydrogen atom splits into triplet $(F=1)$ and singlet $(F=0)$ states. The relative population density of neutral hydrogen atoms in the triplet $(n_1)$ and singlet $(n_0)$ states at temperature $(T)$ can be expressed as $n_1/n_0 = 3\,\exp \left(-T_*/T\right)$, where, $T_* = 68$\,mK corresponds to the temperature of hyperfine transition in neutral hydrogen atoms. Any process that affects the equilibrium temperature, known as the spin temperature $(T_s)$, can alter the relative population density $(n_1/n_0)$. The difference between the spin temperature  $(T_s)$ and CMB temperature $(T_{\gamma})$ over redshift $(z)$ can be expressed in terms of differential brightness temperature $T_{21} = \left(T_s - T_{\gamma}\right)/(1+z)\times \left[1 -\exp(-\tau_{21})\right]$, where $\tau_{21}$ is the optical depth of 21-cm photons \cite{Pritchard:2011xb}.


After recombination, the residual free electrons and CMB photons share the same temperature due to efficient Compton scattering\cite{Pritchard:2011xb}. The free electrons and other plasma species remain in thermal equilibrium due to efficient scattering between them. Therefore, the spin temperature is the same as the CMB temperature, and there will not be any global 21-cm signal \cite{Pritchard:2011xb}. Around the redshift $z\sim 200$, the number density of free electrons becomes very small, and the Compton scattering rate becomes smaller than the Hubble expansion rate. As a result, the gas decouples from CMB radiation and both evolve independently. In the $\Lambda$CDM framework, the temperature CMB and gas fall as $(1+z)$ and $(1+z)^2$, respectively. Hence, the gas or IGM temperature $(T_g)$ falls below the temperature of CMB radiation. During this era, the spin temperature is defined by the IGM temperature via collisional coupling \cite{Pritchard:2011xb}. Therefore, an absorption profile in the global 21-cm signal can be observed. This signal is known as the Dark Ages global 21-cm signal. During this era, the universe is primarily homogeneous and isotropic in the absence of star formation. Hence, the dark ages global 21-cm signal is independent of astrophysical uncertainties.

Observing the Dark Ages 21 cm signal, we need the sensitivity of radio antennas below $45\text{ MHz}$, thus making it extremely difficult to observe from ground-based telescopes in the presence of ionospheric distortion and radio frequency interference (RFI). However, a recently proposed ground-based experiment called MIST might be able to observe the signal in the frequency range of $25-105$ MHz \cite{Monsalve:2023lvo}. Additionally, there are many proposed lunar and space-based experiments, such as FARSIDE \cite{farside}, DAPPER \cite{dapper}, FarView \cite{dapper}, SEAMS \cite{seams}, and LuSee Night \cite{luseenight}, aimed at overcoming Earth's atmospheric distortion and RFI. In Ref. \cite{Okamatsu:2023diy}, authors have shown that the Dark Ages global 21-cm signal's shape is nearly independent of the cosmological parameters. They have shown that using the ratios of brightness temperature at three different redshifts makes it possible to distinguish a standard signal from a signal in the presence of non-standard heating or excess radio background radiation. This ratio holds regardless of the specific values of the cosmological parameters, thus termed as ``dark-ages consistency ratio''.

Studies suggest existence of magnetic fields in galaxy clusters \cite{2002ARA&A..40..319C}, large-scale structures \cite{2018MNRAS.479..776V}, and even in the voids \cite{Neronov:2010gir, Vovk:2011aa}. However, the origin and evolution of large-scale magnetic fields lack a comprehensive understanding. Consequently, numerous studies have investigated the potential origins of these magnetic fields, exploring the hypothesis that they may have originated during the early universe from various physical phenomena. For example, the generation of primordial magnetic fields (PMFs) could be linked to the inflationary era \cite{1988PhRvD..37.2743T, 1992ApJ...391L...1R}, preheating \cite{2001PhRvD..63j3515B}, topological defects \cite{1997MNRAS.287....1S}, the Harrison mechanism \cite{2018CQGra..35o4001H}, and other processes. The generation mechanism determines the specific spectrum of the PMF. For instance, breaking the conformal invariance of electromagnetic fields during inflation can seed primaeval magnetic field, which can further amplify due to ``superadiabatic amplification'' \cite{1988PhRvD..37.2743T}. Considering a power spectrum $P_B(k) \propto k_{l}^{n_B}$, where $n_B$ and $k_l$ are the spectral index and wavenumber, respectively, then the $n_B$ associated with such generation mechanism can take values of $-1$ and $-3$ \cite{1988PhRvD..37.2743T}. Similarly, $n_B$ takes a value of $-2.9$ for a nearly scale-invariant spectrum, where the amplitude of the magnetic field becomes nearly independent of length scale \cite{2005MNRAS.356..778S}. In contrast, $n_B$ can take positive values if the fields are generated from causal processes \cite{Vachaspati:1991nm, PhysRevLett.79.1193, PhysRevD.57.664, Grasso:2000wj, Widrow:2002ud, Giovannini:2003yn, Kandus:2010nw}. In the early universe $(T\sim 0.2\, \rm GeV)$, first-order phase transition can seed a primordial magnetic field with spectral index, $n_B = 0$ \cite{PhysRevLett.51.1488}. In Ref. \cite{Durrer:2003ja}, authors have shown that magnetic fields generated at the electroweak phase transition can take positive spectral index $(n_B = 2)$.

Different constraints on primordial magnetic fields have been obtained from various observations. In Ref. \cite{2017PhRvD..95f3506Z}, authors have used cosmic microwave background temperature polarization to set an upper limit on $\rm B_{1\rm Mpc} < 3.3\,\rm nG$ with 95\% confidence level (CL), and the South Pole Telescope (SPT) placed an upper limit of $\rm B_{1\rm Mpc} < 1.5\,\rm nG$. The combined PLANCK+SPT data tightened this bound to $\rm B_{1\rm Mpc} < 1.2\,\rm nG$ with 95\% CL. The anisotropies in the CMB due to the thermal Sunyaev-Zel'dovich effect in the IGM set an upper bound of $\rm B_{1 \rm Mpc} = 0.1\, nG$ with $n_B = 0.0$ \cite{2017PhRvD..96l3525M}. Similarly, measured $\mu$-type CMB spectral distortion by COBE/FIRAS sets an upper bound on a nearly scale-invariant magnetic field to be $B_n < 40\,\rm nG$, while PIXIE forecasts a tighter bound of $B_n < 0.8\,\rm nG$ \cite{PhysRevLett.85.700, 2014JCAP...01..009K}. The baryon-to-photon number constraints $B_n \lesssim 1.0\,\mu\rm G$ at scales $10^4 < k/h\,\rm Mpc^{-1} < 10^8$ \cite{10.1093/mnrasl/slx195}. Lastly, bounds on the PMFs have also been studied from first-order phase transition \cite{Ellis:2019tjf}, Fermi-LAT \cite{Fermi-LAT:2018jdy}, and TeV blazar \cite{Tavecchio:2010mk}.

The presence of primordial magnetic fields can result in the dissipation of energy into the intergalactic medium (IGM) through magnetohydrodynamic (MHD) effects \cite{PhysRevD.57.3264, PhysRevD.58.083502, 2005MNRAS.356..778S}. The decaying MHD effects, such as ambipolar diffusion and turbulent decay, have the potential to impact the temperature of the IGM \cite{2005MNRAS.356..778S, Chluba:2015lpa, 2019MNRAS.488.2001M}. An efficient magnetic energy dissipation can significantly increase the kinetic temperature of the IGM, potentially raising it above CMB temperature \cite{2005MNRAS.356..778S, Chluba:2015lpa, 2019MNRAS.488.2001M}. Consequently, this process can alter and even erase the global 21-cm signal. Therefore, measuring the global 21-cm signal can constrain primordial magnetic fields \cite{2006MNRAS.372.1060T, Bhatt:2019lwt, Natwariya:2020ksr}.




In this work, we study the effect of primordial magnetic fields on the Dark Ages global 21-cm signal. The $\Lambda$CDM model predicts absorption signals during the dark ages $(z \sim 89)$ and cosmic dawn $(z \sim 17)$. To determine an upper bound on the PMFs, we consider $T_g \leq T_{\gamma}$ at $z = 89$ to ensure that PMFs do not produce an emission signal during the dark ages. Additionally, to avoid an emission signal during cosmic dawn, we include $T_g\leq  T_{\gamma}$ at redshift $z = 17$ as well. These conditions yield the most stringent constraint on the PMFs with spectral indices from $-2.99$ to $-1.58$ compared to all existing and future forecasted constraints. Moreover, the bounds on PMF with $n_B$ values from $-2.84$ to $-1.58$ are derived from the Dark Ages. Therefore, they are free from astrophysical uncertainties as well. The global 21-cm signal from the Dark Ages might provide an accurate insight into the presence of an extra-heating source, such as PMFs. Therefore, we determine the Dark Ages global 21-cm signal in the presence of PMFs. Finally, we demonstrate that the dark-ages consistency ratio can distinguish the presence of extra heating caused by PMFs from the $\Lambda$CDM scenario.

The paper is organized as follows. In Sec. \ref{sec: G21cm}, we describe the evolution of dark ages global 21-cm signal $(T_{21})$ with redshift. In addition, we study the variation of $T_{21}$ due to cosmological parameter uncertainty and the consistency ratio $Q_i$. In Secs. \ref{sec: PMFs} and \ref{sec: Tgas}, we study the thermal evolution of IGM in the presence of primordial magnetic fields via ambipolar diffusion and turbulent decay. We also discuss the evolution of $T_{21}$ in the presence of PMFs. Further in Sec. \ref{sec: Result}, we discuss the constraint on PMFs from dark ages and cosmic dawn $T_{21}$ signal. We also discuss how dark ages global 21-cm signal can put a stringent constraint on PMFs, especially on higher values of the spectral index $(n_B)$. Lastly, in Sec. \ref{conclusion}, we summarize our results with available present and forecast bounds on PMFs. In this work, we have considered a flat $\Lambda$CDM universe with cosmological parameters fixed to $\Omega_b = 0.048, \Omega_m = 0.31,\text{ and } h \equiv H_0/100\, \rm Km/s/Mpc = 0.6766$, where $\Omega_b$ and $\Omega_m$ are the dimensionless energy density parameters for the baryon and matter, and $h$ is the Hubble parameter taken from Planck 2018 \cite{Planck:2018vyg}. 

\section{global 21-cm signal}\label{sec: G21cm}

\subsection{Evolution of global 21-cm signal}\label{sec: Dark_ages_signal}
The intensity distribution of 21-cm photons in CMB can be approximated by Rayleigh-Jeans law. Therefore, the redshifted global (sky-averaged) intensity in terms of temperature with respect to $T_{\gamma}$ can be expressed as

\begin{equation}
	T_{21} = \frac{T_s - T_{\gamma}}{1+z}\left(1 - e^{-\tau_{21}}\right)
\end{equation}
%
In the limit of $\tau_{21} \ll 1$, the above equation can be re-written as \cite{Pritchard:2011xb},

\begin{alignat}{2}
	 T_{21} \approx 27 \left(1 - x_e\right) &\left(1-\frac{T_{\gamma}}{T_s}\right) \left(\frac{1 - Y_p}{1 - 0.24}\right) \sqrt{\left(\frac{0.1424}{\Omega_mh^2}\right)}\nonumber \\ 
  & \times\, \left(\frac{\Omega_bh^2}{0.02242}\right) \sqrt{\left(\frac{1+z}{10}\right)}\,\rm {mK},
	\label{eq: T21}
\end{alignat}
where $Y_p$ represents the helium fraction, and $h$ represents the Hubble parameter in units of $100\mbox{\ }\rm Km/\rm s/\rm Mpc$. In addition to cosmological parameters ($Y_p$, $\Omega_bh^2$ and $\Omega_mh^2$), $T_{21}$ depends upon CMB temperature, spin temperature, and ionization fraction, $x_e=n_e/n_H$. Here, $n_e$ and $n_H$ represent the number density of free electrons and hydrogen, respectively, in the universe.

The spin temperature determines the relative population density of hyperfine states in a HI atom. However, different processes can affect $T_s$ such as Lyman-alpha (Ly$\alpha$) photons via Wouthuysen-Field effect \cite{1952AJ.....57R..31W, 1959ApJ...129..536F}, 21-cm photons from CMB and non-thermal sources via hyperfine transition \cite{Pritchard:2011xb}, and collision between HI atoms and constituents of IGM via spin-exchange \cite{1958PIRE...46..240F}. Therefore, the evolution of $T_s$ can be expressed as,

\begin{equation}
	T_s^{-1} = \frac{T_{\gamma}^{-1}+x_{\alpha}T_{\alpha}^{-1}+x_cT_g^{-1}}{1+x_{\alpha}+x_c}\, .
	\label{eq:spin temp}
\end{equation}
Here, $x_c$, $x_{\alpha}$, and $T_{\alpha}$ represent collisional coupling via spin-exchange, Ly$\alpha$ coupling and effective temperature of Ly$\alpha$ photons, respectively. In the absence of Ly$\alpha$ source during the Dark Ages era--- as there are no star formation during the Dark Ages in the $\Lambda$CDM framework of cosmology, Eq. (\ref{eq:spin temp}) can be written as,

\begin{equation}
	\left(1 - \frac{T_{\gamma}}{T_s}\right) = \frac{x_c}{1+x_c} \left(1 - \frac{T_{\gamma}}{T_g}\right),
	\label{eq:ref_equation}
\end{equation}

\begin{equation}
	x_{c} = \frac{T_{*}}{T_{\gamma}}\frac{n_i\ k_{10}^{iH}}{A_{10}},
\end{equation}
where $T_* = 68 ~\rm mK$ is the temperature equivalent of $21 \mbox{\ }\rm cm$ photons. Here, $n_i$ represents the number density of the species `$i$' present in the IGM while $k_{10}^{iH}$ represents their corresponding collisional spin de-excitation rate. $A_{10} = 2.85\times 10^{-15} ~\rm Hz$ is the Einstein coefficient for spontaneous emission in the hyperfine state. Calculating the de-excitation rate requires quantum mechanical calculations, whose tabulated values corresponding to $k^{HH}_{10}$, $k^{eH}_{10}$, and $k^{pH}_{10}$ can be found in literature \cite{2006nla..conf..296Z, 2007MNRAS.374..547F, 2007MNRAS.379..130F}. The term $n_ik_{10}^{iH}$ can be written as,
\begin{equation}
	n_ik_{10}^{iH} = n_{H} k^{HH}_{10} + n_e k^{eH}_{10} + n_p  k^{pH}_{10},
\end{equation}
where $n_p$ represents the number density of residual protons. The de-excitation rates $k^{HH}_{10}$ , $k^{eH}_{10}$, and $k^{pH}_{10}$ can be approximated in a functional form as follows \cite{2006ApJ...637L...1K, 2001A&A...371..698L, Mittal:2021egv, Pritchard:2011xb},

\begin{alignat}{2}
	k^{HH}_{10} & = 3.1 \times 10^{-17}\left(\frac{T_{g}}{\mathrm{K}}\right)^{0.357}\cdot e^{-32\mathrm{K} / T_g}, \\
	\log_{10}{k^{eH}_{10}} & = -15.607 + \frac{1}{2}\log_{10}\left(\frac{T_g}{\mathrm{K}}\right)\times \nonumber \\ & ~~~~\qquad\qquad \exp{-\dfrac{\left[\log_{10} \left(T_g/\mathrm{K}\right)\right]^{4.5}}{1800}}, \\
	k^{pH}_{10} & = 10^{-16}\Bigg(b_0 + b_1 \log_{10} \left(\frac{T_g}{\mathrm{K}}\right) + b_2 \log^2_{10} \left(\frac{T_g}{\mathrm{K}}\right) \nonumber \\
	& ~~~~\qquad\qquad + b_3\log^3_{10} \left(\frac{T_g}{\mathrm{K}}\right)\Bigg), \label{coupling_coefficients}
\end{alignat}

where, $b_0$, $b_1$, $b_2$, and $b_3$ are equal to $4.28$, $0.24$, $-1.37$, and $0.53$, respectively \cite{ Mittal:2021egv}. All $k^{iH}_{10}$ terms have the dimension of $\rm m^3 \rm s^{-1}$. Here, $k^{iH}_{10}\rm s$ have been approximated under the consideration that $T_g<10^4\, \rm K$. As shown in Fig. \eqref{fig:Gas_temp}, for all the viable cases, the gas temperature remains below $10^4\,\rm K$ during the dark ages; therefore, we can consider the approximation mentioned above. Considering the universe to be electrically neutral $(n_e \approx n_p)$, $(x_c)$ can be written as \cite{Pritchard:2011xb},

\begin{equation}
	x_{c} = \frac{n_HT_{*}}{A_{10}T_{\gamma}}\left\{(1 - x_e)k_{10}^{HH}+ x_e\left(k^{eH}_{10} + k^{pH}_{10}\right)\right\}.   \label{eq10} 
\end{equation}

The IGM and CMB photons remain coupled till the redshift $z\approx 200$, after which Compton scattering $(\Gamma_c)$ starts to become ineffective. Therefore, the gas temperature decouples from the CMB radiation. After $z\approx 200$, the gas temperature evolves adiabatically, i.e. $T_g\propto (1+z)^2 $ and CMB temperature is propotional to $(1+z)$. Hence, the gas temperature falls below the CMB temperature \cite{Pritchard:2011xb}. It can be seen from Eq. \eqref{eq10} that collisional coupling is a function of $x_e$, $T_{\gamma}$, and $T_g$. Thus, $x_c$ reduces from $\sim 50$ at $z \sim 200$ to $\sim 2$ at $z \sim 90$ due to the expansion of the universe. Consequently, $T_s$ remain coupled to $T_g$ till $z\approx 89$, and as soon as $x_c$ become ineffective, $T_s$ start approaching $T_{\gamma}$ causing an expected absorption at redshifts $\left(40 \lesssim z \lesssim 200\right)$.

Around the redshift $z\sim 40$, gas density decreases significantly, and the scattering between hydrogen atoms and other species becomes inefficient. As a result, the spin temperature becomes equal to the CMB temperature, leading to an absence of the global 21-cm signal. Near the redshift of $z\sim 30$,  Lyman alpha photons from the stars strongly couple the IGM temperature to $T_s$, via Wouthuysen-Field effect \cite{1952AJ.....57R..31W, 1959ApJ...129..536F}, leading to absorption in $T_{\gamma}$ at redshifts $(14\lesssim z\lesssim 30)$. Soon after that, the evolution of astrophysical structures heats the IGM, causing $T_g$ greater than $T_{\gamma}$. Consequently, the spin temperature coupled to $T_g$ becomes greater than $T_{\gamma}$. Therefore, we expect an emission profile in the global 21-cm signal until the IGM is completely ionized by the astrophysical sources around $(z\sim 6)$. Furthermore, we expect no global 21-cm signal at redshifts $z\lesssim 6$ in the absence of neutral hydrogen atoms. For a detailed discussion, refer to the review article \cite{Pritchard:2011xb}.

Recently, EDGES have reported an anomalous (flat U-shaped) absorption signal with an amplitude of $-0.5^{+0.3}_{-0.5}$ K at redshifts $14 < z < 20$, centring at $z_{\rm abs} \sim 17$ \cite{Bowman:2018yin}. However, the SARAS 3 experiment has rejected the entire signal with $95\%$ CL \cite{Singh:2021mxo}. Thus, future measurements such as HERA \cite{DeBoer:2016tnn}, REACH \cite{deLeraAcedo:2022kiu}, and MWA-II \cite{Tiwari:2023wzg} might resolve this tension.  This motivated many authors to explain the anomalous signal by invoking different standard and non-standard scenarios. One method is to cool the IGM gas by introducing baryon-dark matter interaction \cite{Barkana:2018qrx, Fialkov:2018xre}. The signal can also be explained by excess-radio background \cite{Lawson:2012zu, Feng:2018rje, Pospelov:2018kdh, Lawson:2018qkc}. The absorption signal in the 21-cm line during cosmic dawn has also been used to constrain various physics, such as sterile neutrino as dark matter \cite{Natwariya:2022xlv}, and energy injection by primordial black holes \cite{Natwariya:2021xki, Saha:2021pqf}.


The Dark Ages signal to be free from astrophysical uncertainties requires accurate modelling of $T_{21}$ for precision cosmology. For instance, the calculation of $k^{iH}_{10}$ terms has been performed in the absence of velocity-dependent non-thermal distribution of hyperfine coupling whose inclusion could lead to a $5\%$ suppression in the dark ages $T_{21}$ signal \cite{2007MNRAS.375.1241H, Pritchard:2011xb}. In addition to the amplitude, the shape of the Dark Ages signal can provide valuable insights. In the following section, we discuss the shape of the Dark Ages global 21-cm signal in detail.

\subsection{Shape of the Dark Ages global 21-cm signal}
The universe was mostly homogeneous and isotropic during the Dark Ages in the absence of star formation. Therefore, the shape of a global 21-cm signal in the frequency range $20\text{ MHz}\lesssim \nu\lesssim 45\text{ MHz}~(30.5\lesssim z \lesssim 70)$ may also provide valuable cosmological information. Recently, authors in Ref. \cite{Okamatsu:2023diy} have pointed out that $T_{21}$ signal's shape varies mildly for different values of the cosmological parameters $\left(\Omega_mh^2, \,\Omega_bh^2, \text{ and, }Y_p\right)$. Therefore, they have introduced a new observable, known as the ``dark ages consistency ratio", capable of testing the consistency of standard cosmology by distinguishing standard $T_{21}$ signal from the presence of any other type of exotic physics. To determine the dependence of $T_{21}$ signal's shape, they have calculated $T_{21}$ value at three different frequencies in the range $20\text{ MHz}\lesssim \nu \lesssim 45\text{ MHz}$. We revisit the same in this section.

\begin{figure}
    \centering
    \includegraphics[width=\linewidth]{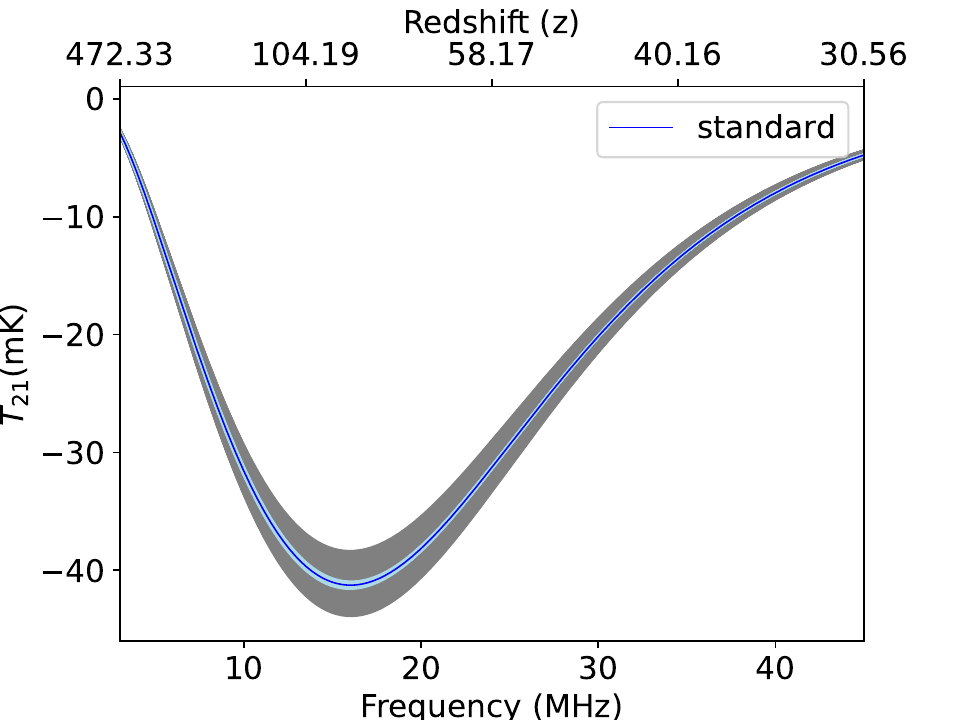}
    \caption{The dark ages $T_{21}$ signals with redshifted frequency upon varying cosmological parameters $(\Omega_bh^2, \Omega_mh^2, Y_p)$ in their corresponding $68\%$ CL range, that is, $(0.02234\pm 0.00014)$, $(0.14205\pm 0.00090)$, and $(0.2436 \pm 0.00395)$ respectively \cite{Planck:2018vyg}, depicted by the grey band. The solid blue line shows $T_{21}$ for the mean values of the cosmological parameters, and the light-blue band depicts variation in $T_{21}^{\prime}$ (Eq. \ref{eq: scaled T21}).}
    \label{fig:subfig1}
\end{figure}

From Eq. (\ref{eq: T21}), it is evident that $T_{21}$ signal's amplitude depends on cosmological parameters: $\Omega_mh^2$, $\Omega_bh^2$, and $Y_p$. Although their values are known accurately, we still vary them within $68\%$ CL, that is, $(0.02234\pm 0.00014)$, $(0.14205\pm 0.00090)$, and $(0.2436 \pm 0.00395)$ respectively, to observe the variations in $T_{21}$ signal \cite{Planck:2018vyg}. In Fig. \eqref{fig:subfig1}, the grey band shows that the amplitude changes evidently. We found a total variation of $\sim 6\,\rm mK$ $\left(-42.5^{+4}_{-2}\,\rm mK\right)$ at $z\simeq89~(\nu\simeq 15.7\text{ MHz})$ . However, the signal's shape changes mildly, especially in the frequency range $20\text{ MHz}\lesssim \nu\lesssim 45 \text { MHz}$. To verify the dependence of the signal's shape on cosmological parameter uncertainties, we scaled $T_{21}$ with $\mathcal{M}(\phi)$. We define $\mathcal{M}(\phi)$ and scaled $T_{21}\,\left(T_{21}^{\prime}\right)$ as follows,
\begin{eqnarray}
T_{21}^{\prime} & = &\frac{T_{21}}{\mathcal{M}(\phi)} \mathcal{M}(\phi^{\prime}),
\label{eq: scaled T21} \\
\mathcal{M}(\phi) & = & \frac{\left[\Omega_bh^2(1 - Y_p)\right]^2}{\sqrt{\Omega_mh^2}}\label{eq: Mphi},\\ \mathcal{M}(\phi^{\prime}) & = & \mathcal{M}(\Omega_m^{\prime}, \Omega_b^{\prime}, h^{\prime}, Y_p^{\prime}).
\end{eqnarray}
The reason for considering $\mathcal{M}(\phi)$ can be easily found from Eq. (\ref{eq: T21}). As discussed earlier, the collisional coupling become ineffective $(\ll 1)$ at redshifts $z\lesssim 69$, and it is proportional to $\left[\Omega_bh^2(1-Y_p)\right]$. As a result at redshifts below $\sim 69$ $\left(\text{or }\nu\gtrsim 20\,\rm MHz\right)$, the brightness temperature can be considered proportional to $\mathcal{M}(\phi)$. The variations in $T_{21}^{\prime}$ are depicted in lightblue band in Fig. \eqref{fig:subfig1}. We find a total variation of $\sim 3\,\rm mK$ $(-42.5^{+2}_{-1}\,\rm mK)$ in $T_{21}^{\prime}$ at redshift $z\sim 89~(\nu\sim 15.7\text{ MHz})$. To quantify the variation in the signal's shape, we randomly select a frequency $33\,\rm MHz ~(z \sim 42)$ as the base frequency and define a ratio $Q_{i}$ as,
\begin{equation}
	{Q}_{i} = \frac{T_{21}(\nu_i)}{T_{21}(33\,\rm MHz)}\, .
	\label{eq: Q}
\end{equation}

\begin{figure}
    \centering
    \includegraphics[width=\linewidth]{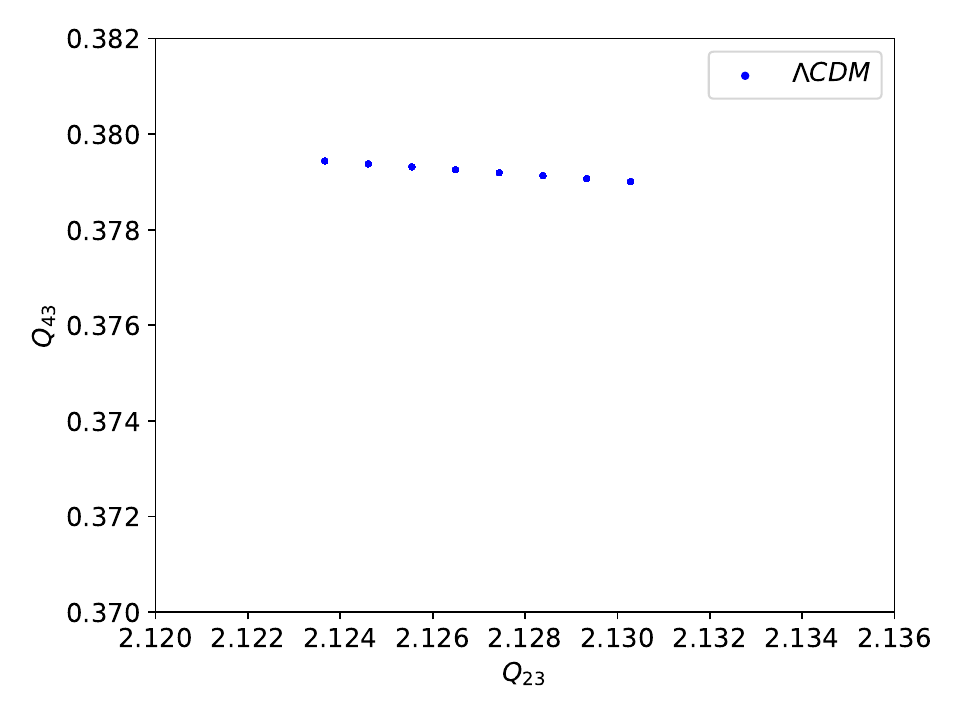}
\caption{Shows variation in $Q_i$ corresponding to the grey band shown in the Fig. \eqref{fig:subfig1}. It can be seen that the ratios $Q_{43}$ attain almost a constant value whereas $Q_{23}$ changes mildly within a range $[2.124, 2.132]$.}
    \label{fig:subfig2}
\end{figure}

In Fig. \eqref{fig:subfig2}, we show the ratios for all the $T_{21}$ signals depicted in the grey band in Fig. \eqref{fig:subfig1}. It can be seen that $Q_{43}$ attains an almost constant value, whereas $Q_{23}$ lies between $[2.124, 2.132]$, evaluted at frequecies $43\,\rm MHz~(z\sim 32)$ and $23\,\rm MHz~(z\sim 60.7)$, respectively. The reason for this can be analyzed from the fact that $Q_{23}$ strongly depends on $\left[x_c/(1+x_c)\right]$, whereas $Q_{43}$ is nearly-independent of $x_c$. Therefore $Q_{43}$ attains a constant value whereas $Q_{23}$ varies mildly for $\nu\gtrsim 20\,\rm MHz~(z\lesssim 69)$. 

Future 21-cm observations may consider any three frequencies between $23\lesssim \nu/\rm MHz\lesssim 45$ to check the consistency of the signal. Thus, any deviation from this value will conclude the presence of non-standard physics during the dark ages. In the following sections, we explore the possibilities of variations in the amplitude and shape of the Dark Ages global 21-cm signal in the presence of PMFs.
\section{Primordial magnetic fields}\label{sec: PMFs}
After the recombination, there are two processes through which the primordial magnetic fields can transfer energy into the IGM--- ambipolar diffusion and turbulent decay. Ambipolar diffusion is important when plasma is partially ionized, whereas non-linear effects can lead to turbulent decay \cite{2005MNRAS.356..778S}. In this work, we follow Sethi and Subramanian's calculation to account for ambipolar diffusion and turbulent decay \cite{2005MNRAS.356..778S}. We consider a statistically isotropic and homogeneous Gaussian random field, whose two-point correlation function in Fourier space is given by \cite{Kunze:2014, 2019MNRAS.488.2001M},
\begin{equation}
	\Big{<} B_i^{\prime}(\mathbf{k})~B_j(\mathbf{k^{\prime}})\Big{>} = \frac{(2\pi)^3\delta_{\rm D}(\mathbf{k - k^{\prime}})}{2}\left(\delta_{ij} - \hat{k}_i \hat{k}_j\right)P_{B}(k),
\end{equation}
where $B_i^{\prime}(\mathbf{k})$ is the Fourier transform of $B_i(\mathbf{x})$ with $\mathbf{k}$ mode. Therefore, the statistical properties of PMFs can be determined by their power spectrum. We consider a power spectrum $(P_{B}(k))$ following a power law in $k$ modes, expressed as \cite{2005MNRAS.356..778S, Kunze:2014, 2019MNRAS.488.2001M}, 

\begin{equation}
	P_B(k) = \frac{B_n^2\ (2\pi)^{n_B+5}}{\Gamma\left[\left(n_B + 3\right)/2\right]}\frac{k^{n_B}}{k_n^{n_B+3}}\, ,
	\label{power_spectrum}
\end{equation}
where, $B_n$ is the PMFs amplitude normalised at a length scale of $1\text{ Mpc}$, $\Gamma$ is the Gamma function, $k_n = 2\pi\text{ Mpc} ^{-1}$, and $n_B$ is the spectral index of PMFs. The magnetic field $B_l^2$ smoothed at any length scale $l$ can be expressed as \cite{2019MNRAS.488.2001M},
\begin{equation}
B_l^2 = \int_{0}^{\infty} \frac{d^3k}{(2\pi)^3} e^{-k^2l^2}P_B(k) = B_n^2\left(k_l/k_n\right)^{n_B+3},
\end{equation}
where $k_l = 2\pi/l$. We note that for $n_B = -3.0$, the magnetic fields become scale-independent. Thus, in this work, we have considered $n_B\gtrsim -3.0$ and $B_n^2 = B_{1\text{Mpc}}^2$. Radiative viscosity can damp the PMFs on small length scales ($k>k_c$) before recombination \cite{PhysRevD.57.3264, PhysRevD.58.083502, 2005MNRAS.356..778S}. Therefore, $P_B(k)$ vanishes on inverse length scales greater than the cut-off scale $(k_{c})$, which can be expressed as follows,

\begin{equation}
	P_B(k) = 0 \quad \forall~k \geq k_{c}\, .
\end{equation}
Previous studies suggest that radiative diffusion damps the primordial magnetic fields on small scales before the recombination epoch, and this damping scale can change over time as the universe evolves \cite{PhysRevD.58.083502}. Therefore, we take the temporal evolution of the cut-off scale as $k_c = g(z)\,k_{c,i}$ with $g(1088) = 1$ \cite{2019MNRAS.488.2001M}, where $k_{c,i}$ is the cut-off scale at $z = 1088$ (end of the recombination epoch) and $g(z)$ determines the temporal evolution of $k_c$. The power spectrum we considered in Eq. \eqref{power_spectrum} holds for $k<k_c$. The $k_{c,i}$ is given as \cite{PhysRevD.58.083502, PhysRevD.65.123004},

\begin{equation}
	k_{c,i}^{-2} = \frac{V_a^2}{\sigma_T}\int_{1088}^{\infty} \frac{(1+z)dz}{n_e(z)H(z)}\,,
	\label{kcut}
\end{equation}
where, $\sigma_T$ is the Thompson scattering cross-section, $H(z)$ is the Hubble expansion rate, $n_e$ is electron number density, and $V_a = B_n(k_c,z)/\sqrt{4\pi \rho_b(z)}$ is Alfvén wave velocity depending on the baryon density $(\rho_b)$. In a matter-dominated universe, the above equation can be approximated as \cite{2019MNRAS.488.2001M},
\begin{equation}
\frac{k_{c,i}}{ k_{n}} \simeq\left[1.32\times 10^{-3} \left( \frac{B_n}{\mathrm {nG}}\right)^2\left(\frac{0.02}{\Omega_bh^2}\right)\sqrt{\frac{\Omega_mh^2}{0.15}}\right]^{-\frac{1}{(n_B+5)}}.
\label{kcut_approx}
\end{equation}

In the presence of a magnetic field, the Lorentz force acts only on the ionized component of the IGM. Thus, it can accelerate the ionized components, thereby increasing the relative velocity between neutral and ionized components. The enhanced relative velocity is damped via ion-neutral collisions in plasma, draining the energy of the magnetic fields \cite{2005MNRAS.356..778S, 10.1093/mnras/116.5.503}. This process is called ambipolar diffusion, which is crucial in a nearly neutral molecular cloud \cite{1992pavi.book.....S}. In the early universe, the neutral component of the IGM contained primarily hydrogen atoms and some fraction of helium atoms. Thus, we ignore the dissipation of energy into helium and formulate the volume rate of energy transfer proportional to the average Lorentz force squared as \cite{1956MNRAS.116..114C, 2005MNRAS.356..778S},
\begin{equation}
	\Gamma_{\rm ambi} = \frac{\rho_{\rm HI}}{\rho_b^2\rho_i} \frac{1}{16\pi^2\gamma_e}\big{|}\left(\vec{\nabla}\times \vec{B}\right)\times \vec{B}\big{|}^2,
	\label{ambi_rate_1}
\end{equation}
where $\rho_{\rm HI}$ and $\rho_i$ are the energy densities of the neutral hydrogen atoms and ions, respectively, present in the plasma. Here, $\gamma_e$ is the ion-neutral coupling coefficient, which can be approximated as $\gamma_e \approx 1.9\times 10^{14}\left(T_g/\rm K\right)^{0.375}\rm cm^3/g/s$ \cite{2005MNRAS.356..778S, Schleicher:2008aa}. We rewrite, $\rho_{\rm HI}/\rho_i\sim (1 - x_e)/x_e$, and $\rho_b\sim m_H n_{H}$, where $m_H$ is mass of hydrogen atom and $n_H$ is represented in comoving coordinate, respectively. We rewrite Eq. (\ref{ambi_rate_1}) as \cite{Natwariya:2020ksr},
\begin{equation}
	\Gamma_{\rm ambi} = \frac{(1 - x_e)/x_e}{16\pi^2\gamma_e(m_Hn_H)^2}\big{|}\left(\vec{\nabla}\times \vec{B}\right)\times \vec{B}\big{|}^2.
	\label{ambipolar_rate_eqn}
\end{equation}
On substituting Eq. \eqref{power_spectrum} with $g(z)$, we can calculate $\vec{L}= (\vec{\nabla}\times \vec{B})\times \vec{B}$ at redshift $z$ as \cite{Kunze:2014, 2019MNRAS.488.2001M},
\begin{equation}
	\big{|}\vec{L}\big{|}^2 = \int\int \frac{d^3k_1d^3k_2}{(2\pi)^6}k_1^2P_B(k_1)P_B(k_2)g(z) ^{2n_B+8}(1+z)^{10}
	\label{Lorentz_Force}
\end{equation}
It should be noted that the derivative operator in the Lorentz force $\vec{L}$ is taken with respect to proper coordinate and $B \sim B_n(1+z)^2$, where $B_n$ is the present-day magnetic field strength, thus $\vec{L}\propto (1+z)^5$. 

Before recombination, turbulent motion in plasma is heavily damped due to large radiative viscosity. However, the damping lowers down after recombination, which leads to the transfer of energy to smaller scales (smaller than magnetic Jeans length) from non-linear interactions, thus dissipating the magnetic field on larger scales--- known as turbulence decay \cite{Kunze:2014}. In the matter-dominated epoch, the approximate decay rate for a non-helical field is given by \cite{2005MNRAS.356..778S},
\begin{equation}
	\Gamma_{\rm decay} = \frac{3\,q_B\,E_B\,H(z)}{2}\frac{[\ln(1+t_d/t_{\rm rec})]^{q_B}}{\left[\ln(1+t_d/t_{\rm rec}) + \ln(t/t_{\rm rec})\right]^{1+q_B}}
	\label{turb_decay_eqn}
\end{equation}
where, $q_B = 2(n_B+3)/(n_B+5)$, $t_d = 1/\left(k_cV_a\right)$ for $n_B>-3.0$ represents Alfvén time scale for cut-off scale \cite{2005MNRAS.356..778S}, $t_{\rm rec} = 2/(3H)$ is recombination time-period, and $E_B = B^2/8\pi$ is the magnetic energy density. We represent $E_B$ with $g(z)$ as \cite{2019MNRAS.488.2001M},
\begin{equation}
	E_B = \frac{1}{8\pi}\int \frac{d^3k}{(2\pi)^3}P_B(k)g(z)^{n_B+3} (1+z)^{4}.	
	\label{mag_enrg}
\end{equation}
The evolution of $E_B$ with redshift in the presence of energy dissipation from Eqs. \eqref{ambipolar_rate_eqn} and \eqref{turb_decay_eqn} can be expressed as \cite{2005MNRAS.356..778S, Kunze:2014, 2019MNRAS.488.2001M},
\begin{equation}
	\frac{dE_B}{dz} = 4\frac{E_B}{1+z} + \frac{1}{(1+z)H(z)}\left(\Gamma_{\rm decay} + \Gamma_{\rm ambi}\right).
	\label{mag_enrg_cons}
\end{equation}
Here, the first term corresponds to the redshifting of magnetic energy due to the adiabatic expansion of the universe. The second term accounts for the energy dissipated at length scales smaller than magnetic Jeans length via turbulent decay \cite{2005MNRAS.356..778S, Chluba:2015lpa, 2019MNRAS.488.2001M}. The third term accounts for the energy lost from the magnetic fields to increase the relative velocity of the ionized component of IGM. These ionized components heat the IGM via collisions with the neutral components \cite{2005MNRAS.356..778S, Chluba:2015lpa, 2019MNRAS.488.2001M}. We then rewrite Eqs. \eqref{ambipolar_rate_eqn}, \eqref{turb_decay_eqn} and \eqref{mag_enrg} in term $k_c$ and substitute in the above equation to determine $g(z)$ evolution as,
\begin{equation}
        \frac{dg}{dz} = \frac{8\pi^3 k_c^{-(n_B+3)}g^{-(n_B+2)}}{(1+z)^5H(z)}\left[\Gamma_{\rm ambi} + \Gamma_{\rm decay}\right]\Big{|}_{k = k_c}\, .
        \label{eq:g(z)}
\end{equation}
\section{Thermal evolution of IGM in the presence of PMF}
\label{sec: Tgas}

The evolution of the IGM temperature with redshift in the absence of any exotic source of energy injection can be expressed as follows \cite{Peebles:1968ja, Seager:1999bc, Chluba:2015lpa}, 
\begin{equation}
    \frac{dT_g}{dz} = 2\frac{T_g}{1+z} + \frac{\Gamma_c}{(1+z)H} \left(T_g - T_{\gamma}\right),
    \label{Gas_Evolution}
\end{equation}
where the first and second terms on the right-hand side represent the adiabatic cooling of the IGM and the coupling between CMB and IGM due to Compton scattering, respectively \cite{1965PhFl....8.2112W}. The Compton scattering rate is defined as,
\begin{equation*}  
    \Gamma_c = \frac{8 n_e\sigma_T a_rT_{\gamma}^4 (z)}{3m_e n_{\text{tot}}} ,\label{Compton_scattering}
\end{equation*}
where $m_e$, $\sigma_T$, and $f_{He} = 0.08$ represent the rest mass of an electron, Thomson scattering cross-section, and helium fraction, respectively. Whereas, $a_r = 7.57\times 10^{-16}$~J\,$\text{m}^{-3}\,\text{K}^{-4}$ represents the radiation density constant, $n_{\rm tot} = n_H(1+f_{He}+x_e)$ represents the total number density of gas, and $T_{\gamma}(z) = T_{0}(1+z)$, $T_0 = 2.725\rm K$ \cite{Seager:1999bc, Seager:1999km}. The evolution of the ionization fraction can be expressed as \cite{Peebles:1968ja, Seager:1999bc, Chluba:2015lpa},
\begin{equation}
	\frac{dx_e}{dz} = \frac{\mathcal{P}}{(1+z)H}\left[n_Hx_e^2\alpha_B - (1-x_e)\beta_Be^{-E_{\alpha}/T_{\gamma}}\right],
	\label{xe_evolution}
\end{equation}

here $\mathcal{P}$ is Peebles coefficient, while $\alpha_B$ and $\beta_B$ are the case-B recombination and photo-ionization rates, respectively \citep{Seager:1999bc, Seager:1999km, Mitridate:2018iag}. The Peebles coefficient is given by \cite{Peebles:1968ja, DAmico:2018sxd},
\begin{equation*}
    \mathcal{P} = \frac{1+ \mathcal{K}_H\Lambda_Hn_H(1-x_e)}{1+ \mathcal{K}_H(\Lambda_H+\beta_H)n_H(1-x_e)},
\label{peeble_coefficient}
\end{equation*}
where $\mathcal{K}_H = \pi^2/(E_{\alpha}^3H)$, $E_\alpha = 10.2\,\rm eV$, and $\Lambda_H = 8.22/\text{sec}$ represents redshifting Ly${\alpha}$ photons, rest frame energy of Ly$\alpha$ photon, and 2S-1S level two-photon decay rate in hydrogen atom respectively \cite{PhysRevA.30.1175}. 

As the star formation takes place, their Ly-$\alpha$ and X-ray radiations start to heat the gas \cite{Furlanetto:2006jb}. The radiation can modify the thermal and ionization evolution of the IGM significantly after cosmic dawn. For a detailed review, please refer to the articles \cite{Furlanetto:2006jb, Pritchard:2011xb}. To include these effects, we follow the articles \cite{Harker:2015uma, Harker:2011et, Mirocha:2015jra}.
%
%
%
%
In these articles, the authors have used the Markov Chain Monte Carlo (MCMC) technique to extract a global 21cm signal in the presence of foreground signals and parametrize the global 21-cm signal using a series of $tanh$ functions with three free parameters. The three free parameters are $A_{(i, \rm ref)}$, ${z_{i,0}}$, and $\delta z_{i}$, where they represent the step height, pivot redshift, and duration, respectively. The $tanh$ approach can mimic the shape of typical global 21-cm signal models extremely well and can be immediately related to the physical properties of the IGM \cite{Mirocha:2015jra}. The $tanh$ parameterization is defined as \cite{Harker:2015uma, Mirocha:2015jra, PhysRevD.98.103529},
\begin{equation}
    A_i = A_{(i,\rm ref)}\left(1+ \tanh \left[\frac{z_{0i} - z}{dz_{i}}\right]\right).
\end{equation}
The ionization fraction and X-ray heating due to star formation are defined by $A_{xe}$ and $A_{X}$, respectively \cite{PhysRevD.98.103529}. The free parameters and their fiducial values associated with $A_{xe}$ are $\{A_{(xe,\rm ref)}, z_{xe,0}\,\text{and}\,\delta z_{xe}\}$ and $\{1, 9, 3\}$, respectively. Similarly, the parameters and fiducial values for $A_{X}$ are $\{A_{(X,\rm ref)}, z_{X}\,\text{and}\,\delta z_{X}\}$ and $\{1000\,\textrm{K}, 12.75, 1\}$, respectively \cite{PhysRevD.98.103529}. To include the ionization fraction, we add $A_{xe}$ to Eq. \eqref{xe_evolution}. Whereas, X-ray heating is incorporated by adding $(dA_{X}/dz)$ to Eq. \eqref{Gas_Evolution} \cite{PhysRevD.98.103529}. In the presence of PMFs and X-ray heating, Eq. \eqref{Gas_Evolution} and Eq. \eqref{xe_evolution} can be written as follows \cite{2005MNRAS.356..778S, Schleicher:2008aa, Chluba:2015lpa, 2019MNRAS.488.2001M, PhysRevD.98.103529},

\begin{alignat}{2}
	\frac{dT_g}{dz} = ~&\frac{dT_g}{dz}\bigg{|}_{\rm Eq. \eqref{Gas_Evolution}} +  \frac{dA_{X}}{dz} \nonumber \\ & -\frac{2}{3n_{\rm tot}(1+z)H(z)} \left(\Gamma_{\rm ambi} + \Gamma_{\rm decay}\right),
	\label{Gas_Evolution_full_eqn}\\
    \frac{dx_e}{dz} = ~&\frac{dx_e}{dz}\bigg{|}_{\rm Eq. \eqref{xe_evolution}} + A_{xe}. \label{xe_Evolution_full_eqn}
\end{alignat}

In Eq. \eqref{Gas_Evolution_full_eqn}, the last term on the right-hand side represents the magnetic heating of the IGM.  
The primordial magnetic field cannot ionize the IGM directly. Instead, it can heat the IGM by dissipating energy via ambipolar diffusion and turbulent decay. At high temperatures, residual electrons can efficiently collide with neutral hydrogen atoms to ionize the IGM \cite{Chluba:2015lpa}. However, the collisional ionization is exponentially suppressed by $\exp(-|E_{1s}/T_g|)$, where $E_{1s} = -13.6\,\rm eV$. Therefore, it is effective for kinetic gas temperature greater than $10^4\,\rm K$ \cite{Chluba:2015lpa}. In the present work, all the possible cases keep $T_g\ll 10^4\,\rm K$ (Fig. \ref{fig:Gas_temp}); therefore, we have ignored this effect. In the below section, we determine the thermal evolution of IGM and dark ages global 21-cm signal in the presence of PMFs.

\section{Results and Discussions} 
\label{sec: Result}

\begin{figure}[htbp]
    \centering
    \subfigure[]{
        \includegraphics[width=0.45\textwidth]{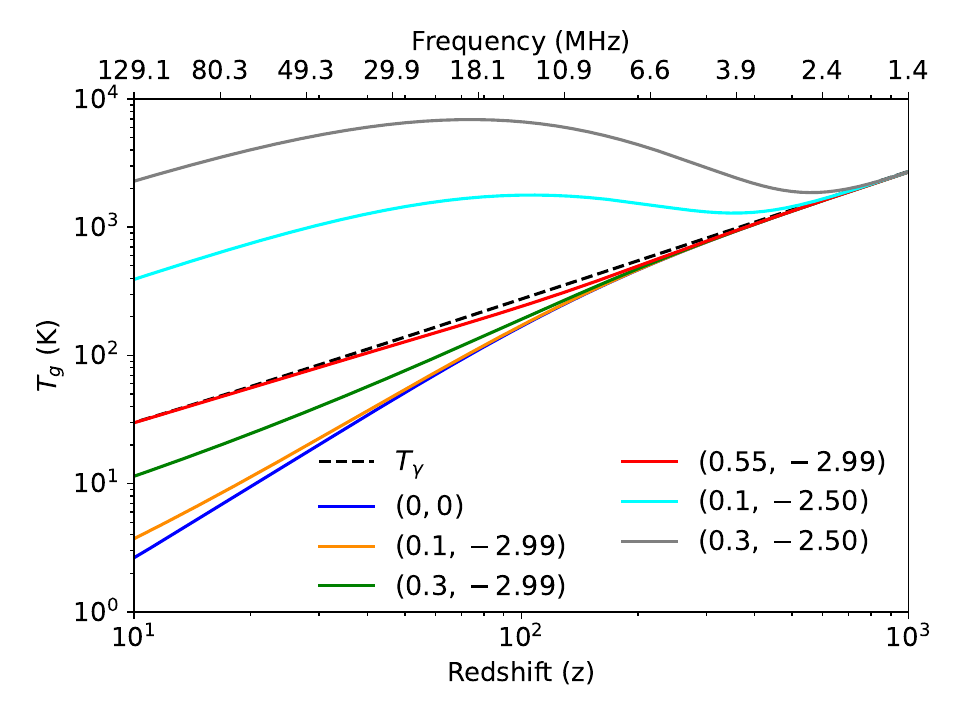}
        \label{fig:Gas_temp_a}
    }
    \subfigure[]{
        \includegraphics[width=0.45\textwidth]{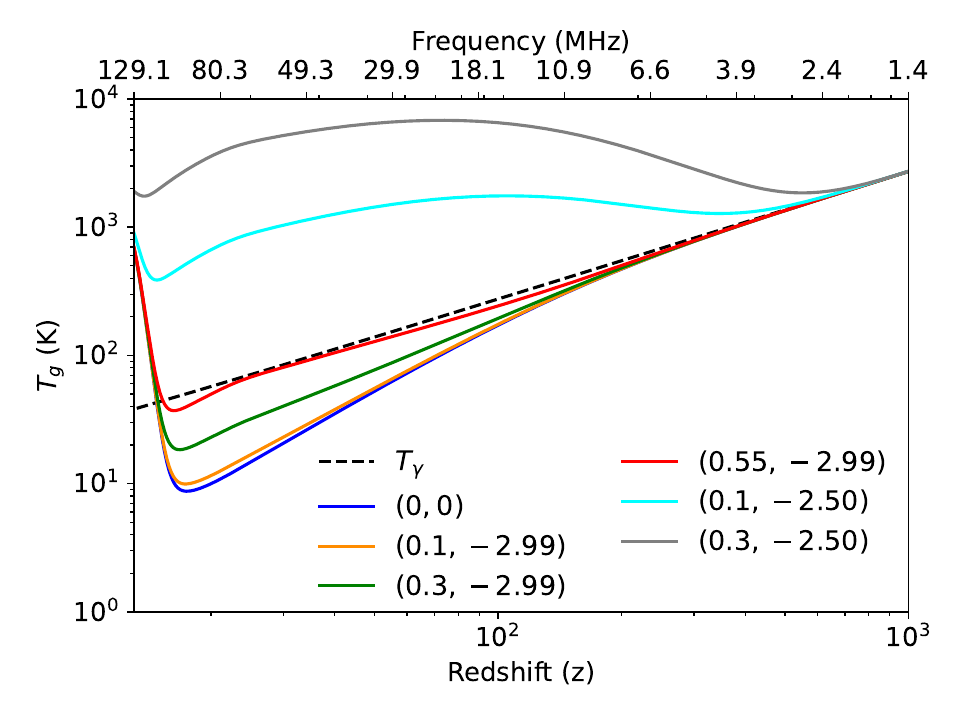}
        \label{fig:Gas_temp_b}
    }
    \caption{These figures represent thermal evolution IGM with redshift. The black dashed line represents CMB temperature, and the blue solid line represents standard IGM temperature evolution. \emph{Top:}
    The IGM evolution due to dissipation of magnetic energy for different $(B_n/\text{nG}, n_B)$ values in the absence of X-ray heating. \emph{Bottom:} Thermal evolution of IGM in the presence of both X-ray and PMF heating (Eq. \ref{Gas_Evolution_full_eqn}).}
    \label{fig:Gas_temp}
\end{figure}

In Fig. \ref{fig:Gas_temp_a}, we have shown 
 the thermal evolution of IGM in the presence of primordial magnetic fields without the X-ray heating effect. We simultaneously solved Eqs. \eqref{eq:g(z)}, \eqref{xe_evolution}, and \eqref{Gas_Evolution_full_eqn}  with the initial conditions:  $g(1088) = 1$ \cite{2019MNRAS.488.2001M}, $T_g = 2967.6\text{ K}$, and $x_e = 0.1315$ at redshift $z = 1088$ taken from \texttt{Recfast++} \cite{10.1111/j.1365-2966.2010.16940.x, 10.1111/j.1365-2966.2010.17940.x}. 
The evolution of CMB temperature is represented by black dashed. In the absence of PMF, the evolution of gas temperature is shown by the blue solid lines. To examine the effects of the magnetic heating on IGM, we increase the strength of $B_n$ while keeping the spectral index $(n_B)$ fixed to $-2.99$. We observe an increase in the IGM temperature, eventually reaching $T_\gamma$--- depicted by the solid orange, green, and red lines, where we increased $B_n$ from $0.1-0.55\,\rm nG$, respectively. We note that, despite increasing the magnetic field strength $(B_n)$, the redshift of decoupling remained almost unaffected. However, increasing $n_B$ from $-2.99$ to $-2.50$ results in an early heating of IGM. This effect is shown by the cyan and grey solid lines, where $n_B$ is $-2.50$ while $B_n$ are $0.1\,\rm nG$ and $0.3\,\rm nG$, respectively. Compared to the orange and green solid lines, where $n_B = -2.99$, significant heating of IGM due to an increase in the $n_B$ value is observed.
%

The behaviour of magnetic heating on IGM can be interpreted by analysing Eq. \eqref{mag_enrg_cons}. The redshift dependence of $\Gamma_{\rm ambi}$ is proportional to $(1+z)^{3.625}$ for $z\gtrsim 200$, and proportional to $(1+z)^{3.25}$ otherwise. Furthermore, by analysing Eqs. \eqref{power_spectrum} and \eqref{Lorentz_Force}, we find the dependence of $\Gamma_{\rm ambi}$ on $n_B$ and $B_n$ to be $\left(\Gamma\left[(n_B+3)/2\right]\right)^{-2}$ and $B_n^4$, respectively. From Eq. \eqref{turb_decay_eqn}, it can be seen that upon ignoring the logarithmic terms, the variation of $\Gamma_{\rm decay}$ with redshift follows $(1+z)^{5.5}$. Similarly from Eq. (\ref{mag_enrg}), the dependence of the magnetic energy density on $n_B$ and $B_n$ can be found to be $\left(\Gamma\left[(n_B+3)/2\right]\right)^{-1}$ and $B_n^2$, respectively. Thus, for a fixed $n_B$ and $B_n$, the magnetic heating of IGM is dominated by turbulent decay at high redshifts compared to the ambipolar diffusion, which dominates the lower redshift regime. Additionally, the quantified dependency of $\Gamma_{\rm decay}$ and $\Gamma_{\rm ambi}$ on $n_B$ for a fixed $B_n$ and $z$ can be analysed by calculating $\Gamma[(n_B+3)/2]$. For example, $(\Gamma[(n_B+3)/2])^{-2}$ takes a value of $2.5\times 10^{-5}$, $9.3\times 10^{-4}$, and $0.076$ for $n_B$ equals to $-2.99$, $-2.94$, and $-2.5$, respectively. Therefore, a small change in $n_B$ can lead to a huge difference in the dissipation of magnetic energy. Larger values of $n_B$ enhance the energy dissipation rate, resulting in a rapid draining of magnetic energy at higher redshifts. Hence, the magnetic heating gradually becomes ineffective, and $T_g$ starts decreasing after a certain redshift. In Fig. \ref{fig:Gas_temp_a}, we present this in the cyan and grey solid lines, where we have fixed $n_B$ to $-2.50$ while taking $B_n$ as $0.1\,\rm nG$ and $0.3\,\rm nG$, respectively. We find that both dissipate magnetic energy efficiently into the IGM till the redshift $\sim 122$ and $\sim 85$, respectively. However, soon after that, $\Gamma_{\rm ambi}$ and $\Gamma_{\rm decay}$ became ineffective--- leading to a fall in $T_g$.

The cosmic dawn era is driven by star formation history. Therefore, to study the evolution of $T_g$ in the presence of magnetic heating as well as X-ray heatings from star formation, we simultaneously solve Eqs. \eqref{eq:g(z)}, \eqref{xe_Evolution_full_eqn}, and \eqref{Gas_Evolution_full_eqn} with the aforementioned initial conditions. In Fig. \ref{fig:Gas_temp_b}, the blue solid line shows an increase in $T_g$ at redshifts $(z \lesssim 25)$ resulting from X-ray heating in the absence of PMF. The gas temperature increases when we include the magnetic field. For example, $T_g$ increased by approximately $1\,\rm K$ at $z=17$ in the presence of magnetic field with $(B_n/\rm nG, n_B)$ equal to $(0.1, -2.99)$--- shown in the orange solid line. However, on increasing $B_n/\rm nG$, we notice a mild decrease in $T_g$ at redshifts $z\lesssim 25$. For instance, $T_g$ takes a value of $18.35\,\rm K$ at $z = 17$ for $B_n = 0.3\, \rm nG$--- shown in green solid line. Whereas the gas temperature increases to $20.24\,\rm K$ at $z = 17$ for the same magnetic field in the absence of X-ray heating (Fig. \ref{fig:Gas_temp_a}). The reason for such decrement can be interpreted by analyzing ambipolar diffusion (Eq. \ref{ambipolar_rate_eqn}). 
The energy dissipation via ambipolar diffusion is proportional to $(1-x_e)/x_e$. Therefore, $\Gamma_{\rm ambi}$ become subdominant as the X-ray photons increase the ionization fraction at redshifts $(z \lesssim 25)$. This leads to a fall in the gas temperature until X-ray heating dominates. We compare the evolution of $T_g$ for $(B_n/\rm nG, n_B)$ equals to $(0.55, -2.99)$ in the absence and presence of X-ray heating, depicted in the red solid line in Fig. \ref{fig:Gas_temp_a} and \ref{fig:Gas_temp_b}, respectively. We find that $T_g$ approaches $T_{\gamma}$ in the absence of X-ray heating at redshifts $(z \lesssim 30)$. Whereas, in the presence of X-ray heating, $T_g$ approaches $T_{\gamma}$ at $z = 30$ but decreases thereafter until X-ray heating becomes dominant.
\begin{figure}[htbp]
    \centering
    \includegraphics[width=\linewidth]{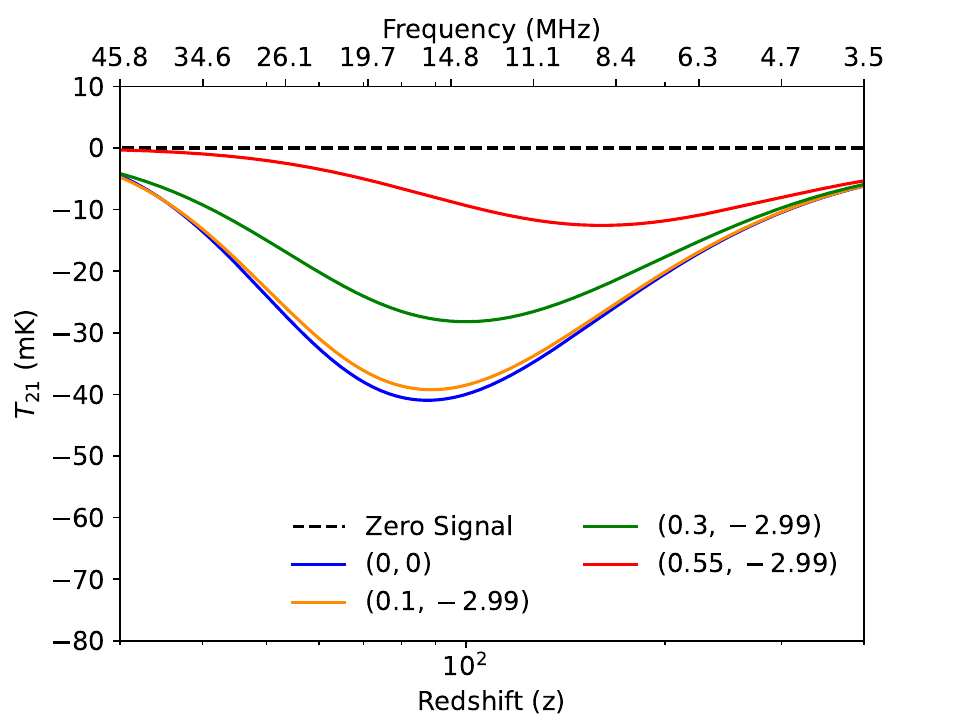}    
	\caption{Dark ages global 21-cm signal with redshift in the presence and absence of PMFs. The blue solid line depicts the standard global 21-cm signal. The black dashed line represents zero or non-existence of a signal, whereas other solid lines represent the presence of PMFs for different $(B_n/\text{nG}, n_B)$ values.}\label{fig:T21_plot}
\end{figure}

In Fig. (\ref{fig:T21_plot}), we have shown the effects of PMFs on dark ages $T_{21}$ signal by evaluating Eq. (\ref{eq: T21}). In Fig. (\ref{fig:T21_plot}), the black dashed line depicts $T_{21} = 0\,\rm mK$, which means the absence of dark ages global 21-cm signal. The blue solid line represents $T_{21}$ signal in the absence of PMFs. We fixed $n_B$ to $-2.99$ and plotted $T_{21}$ signal for $B_n/\rm nG$ equal to $0.1$ and $0.3$, in orange and green solid lines repsectively. Furthermore, the $T_{21}$ signal depth reduce to $\sim 13\,\rm mK$ on increasing $B_n/\rm nG$ to $0.55$--- shown in red sold line. We find that, as the gas temperature increases due to magnetic heating, the depth of $T_{21}$ becomes shallower. This occurs because the ratio $(T_{\gamma}/T_g)$ tends toward unity. Moreover, we find that on increasing magnetic heating, the peak position of $T_{21}$ shifted to higher redshifts. In the below paragraph, we will discuss the detection possibilities of the dark ages global 21-cm signal.

%
\begin{figure}[htbp]
    \centering
    \includegraphics[width=\linewidth]{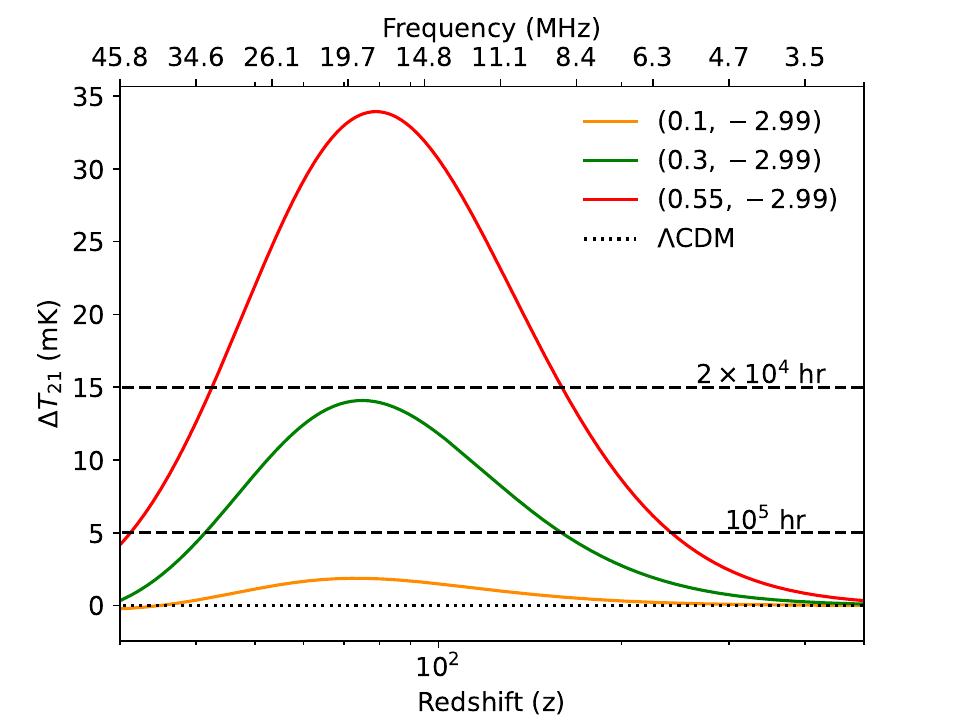}
    \caption{Shows the difference between the 21-cm signal for the $\Lambda$CDM model and the scenarios including primordial magnetic fields $\left(\Delta T_{21} = T_{21}^{\Lambda\rm CDM}-T_{21}^{\rm PMF}\right)$. The black-dashed line represents uncertainties in dark ages $T_{21}$ signal with integration time of observation. For an integration time of 20,000 and $10^5$ hours, the uncertainty in the detection of the standard $T_{21}$ signal becomes 15 mK and 5 mK, respectively \cite{Burns:2020gfh, Rapetti:2019lmf}. Thus, the green and red curves can be measured with an integration time of less than 20,000 hours which represents PMFs with spectral index fixed at $-2.99$ while $B_n/\text{nG}$ equals 0.3 and 0.55 respectively.}
    \label{fig:deltaT21_plot}
\end{figure}

Future lunar-based experiments are expected to detect the dark ages $T_{21}$ signal at redshift $z \sim 89$ with a $68\%$ confidence level $(\Delta T_{21} \sim 5\text{ mK})$ over an integration time of $10^5$ hours \cite{Burns:2020gfh, Rapetti:2019lmf}. Here, $\Delta T_{21}$ represents the deviation of the detected $T_{21}$ from the standard dark ages signal, that is, $\Delta T_{21} = T_{21}^{\Lambda\rm CDM}-T_{21}^{\rm PMF}$. In Fig. \eqref{fig:subfig1}, we have shown that uncertainty in cosmological parameters can cause variations in the $T_{21}$ signal depth of about $6\,\rm mK$--- depicted by the grey band. The mean value of $T_{21}$ at $z= 89$ $(15.78\,\rm MHz)$ is $-42\,\rm mK$, with upper and lower values of $-38\,\rm mK$ and $-44\,\rm mK$, respectively. Thus, any observation resulting in $T_{21}$ outside this grey band would suggest new physics beyond the standard cosmological model. We require the observations to have an accuracy better than $\sim 6\,\rm mK$ to confirm the existence of physics beyond $\Lambda$CDM. In Fig. \eqref{fig:deltaT21_plot}, we plot $\Delta T_{21}$ in the presence of magnetic fields. The black-dotted line corresponds to $\Delta T_{21}=0$ representing the $\Lambda\rm CDM$ signal. Furthermore, the two black-dashed lines show $\Delta T_{21}$ equal to $15\,\rm mK$ and $5\,\rm mK$ for integration times of 20,000 hours and $10^5$ hours, respectively \cite{Burns:2020gfh, Rapetti:2019lmf}. From Fig. \eqref{fig:deltaT21_plot}, we find that detecting the existence of a primordial magnetic field with $(B_n/\rm nG, n_B)$ equal to $(0.1\,\rm nG, -2.99)$, as represented by the orange solid line, requires a well-calibrated instrument with detection uncertainties less than $5\,\rm mK$. In contrast, an integration time of $10^5$ hours could detect the existence of magnetic field $(B_n/\rm nG, n_B)$ equal to $(0.3\text{ nG}, -2.99)$--- shown in the green solid line. Whereas, detecting PMF with $(0.55\text{ nG}, -2.99)$, shown by the red solid line, is possible with an integration time of 20,000 hours. Therefore, measuring the Dark Ages signal could provide an upper bound on magnetic field strengths and spectral indices.

\begin{figure}[htbp]
    \centering
    \includegraphics[width=\linewidth]{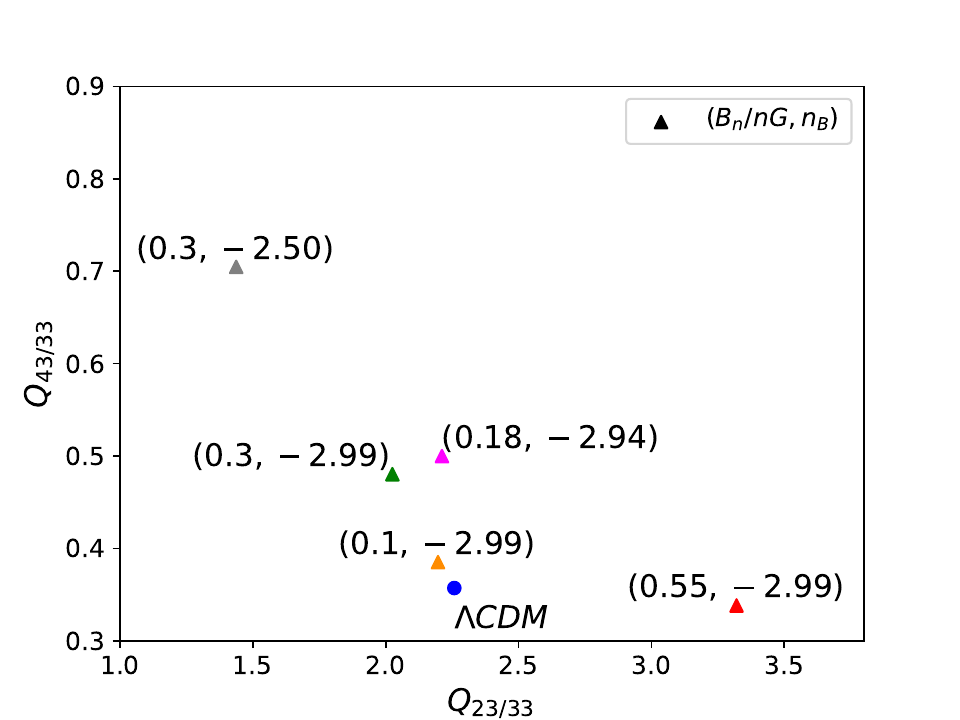}
    \caption{Shows the consistency ratio evaluated from Eq. \ref{eq: Q} for different $(B_n/\text{nG}, n_B)$ values. The $Q_{43}$ and $Q_{23}$ values are calculated at base frequency $\nu_i = 33\text{ MHz}~(z = 42)$ for reference frequencies $43\,\rm MHz$ $(z \sim 32)$ and $23\,\rm MHz$ $(z \sim 60.7)$.}
    \label{fig:consistency_ratio}
\end{figure}

In Fig. (\ref{fig:consistency_ratio}), we present the consistency ratios in the presence of magnetic fields. The blue dot represents $Q_{i/33}$ for $\Lambda$CDM scenario at reference frequencies $43\,\rm MHz$ $(z \sim 32)$ and $23\,\rm MHz$ $(z \sim 60.7)$. Whereas all other $Q_{i/33}$ ratios are calculated in the presence of magnetic fields with different $(B_n/\rm nG, n_B)$ values. We find that all the $Q_{i/33}$ ratios have different values compared to the $\Lambda$CDM scenario. We varied the base frequency $(33\,\rm MHz)$ and reference frequencies and have verified that, for all the $Q_{i}$ in the frequency range $20-45\,\rm MHz$ acquires different values distinguishing PMFs cases from $\Lambda\rm CDM$. Therefore, as suggested by the authors in Ref. \cite{Okamatsu:2023diy}, the ``dark-ages consistency ratio'' might be useful for future experiments to check the consistency in their early stages of observations. However, we have not explored the detection aspect of the ratio in this work.

\begin{figure}
    \centering
    \includegraphics[width=\linewidth]{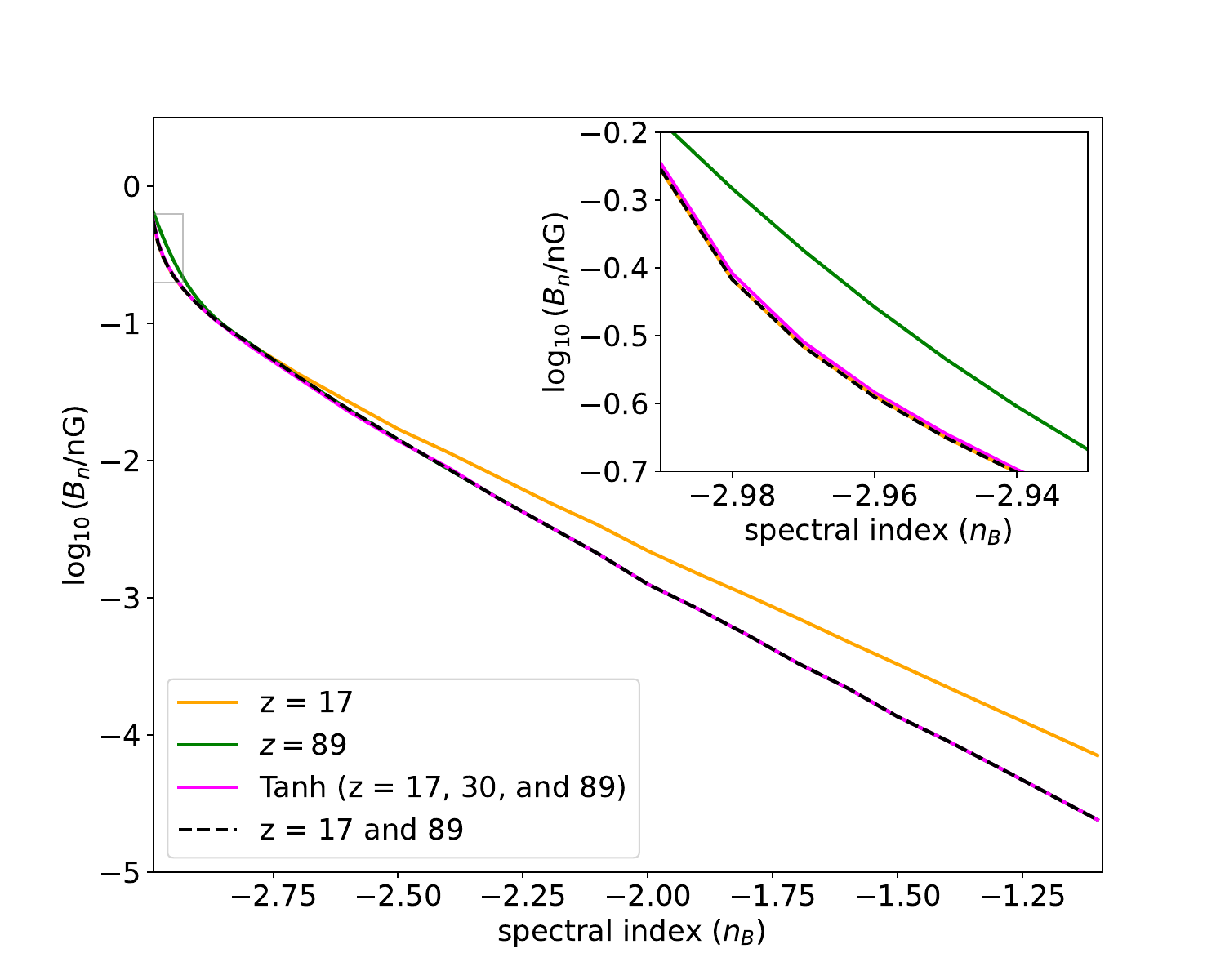}
    \caption{Shows maximum values of $B_n$ to the corresponding spectral indices, considering $T_g\leq T_{\gamma}$ for different redshifts. The bounds presented in the solid orange and green are evaluated at $z= 17$ and $z= 89$, respectively. Whereas the bounds presented in the black dashed line are evaluated at $z= (17,89)$ simultaneously. The upper bounds on $B_n$ in the presence of X-ray heating are shown in magenta solid lines, which are calculated at redshifts $z = (17,30,89)$ simultaneously.}
    \label{fig:constrain_plot_a}
\end{figure}

We then derive upper bounds on the primordial magnetic fields from the global 21-cm signal.
To obtain upper bounds on the present-day strength and spectral index of PMF, we consider $T_g \leq T_{\gamma}$ during the Dark Ages and the cosmic dawn era. First, we consider $T_g \leq T_{\gamma}$ at $z = 89$ to restrict an emission signal during the dark ages. In Fig. \eqref{fig:constrain_plot_a}, we plot the bounds derived from this condition in the green solid line. We then evaluate bounds from the cosmic dawn era by restricting emission at $z=17$--- shown in the orange solid line. We find that the Dark Ages signal can set stronger bounds on magnetic fields with spectral index $-2.84 \leq n_B \leq -1.0$ than the cosmic dawn signal. On the contrary, the cosmic dawn signal sets comparatively stronger bounds on $-2.99 \leq n_B \leq -2.85$ spectra. For instance, the dark ages upper bound on $B_n/\rm nG$ for $n_B$ equal to $-2.99$ and $-2.95$ are $0.65$ and $0.29$, respectively. Whereas $B_n/\rm nG$ takes the values of $0.55$ and $0.22$, respectively, when obtained from the cosmic dawn era. The magnetic fields with greater spectral index dissipate energy effectively at higher redshifts. Therefore, the Dark Ages provide stronger upper bounds on $B_n$ for higher $n_B$ values. Furthermore, to restrict emission signals during the cosmic dawn and the Dark Ages era, we consider $z=17$ and $z=89$ together and obtain the bounds--- shown in the black dashed line. The bounds extracted from the cosmic dawn era depend on astrophysical uncertainties. Therefore, we re-derived the bounds considering $T_g \leq T_{\gamma}$ at redshifts $17$, $30$, and $89$ in the presence of X-ray heating. As we have explained earlier, $T_g$ decreases at redshifts $z\lesssim 25$ in the presence of X-ray heating. Therefore, to restrict an emission signal between $z=17$ and $z = 89$, we considered $z=30$. Subsequently, the maximum values of $B_n$ for $-2.99 \leq n_B \leq -2.85$ shown in the magenta line are greater compared to the black-dashed and orange lines. Moreover, we can consider different astrophysical models where $x_e$ values can increase more during the cosmic dawn era than in the current scenario. Then, the maximum values of $B_n$ obtained from the cosmic dawn era will also increase. We leave the detailed study of primordial magnetic fields with star formation for future work.

\begin{figure}[htbp]
    \centering
    \includegraphics[width=\linewidth]{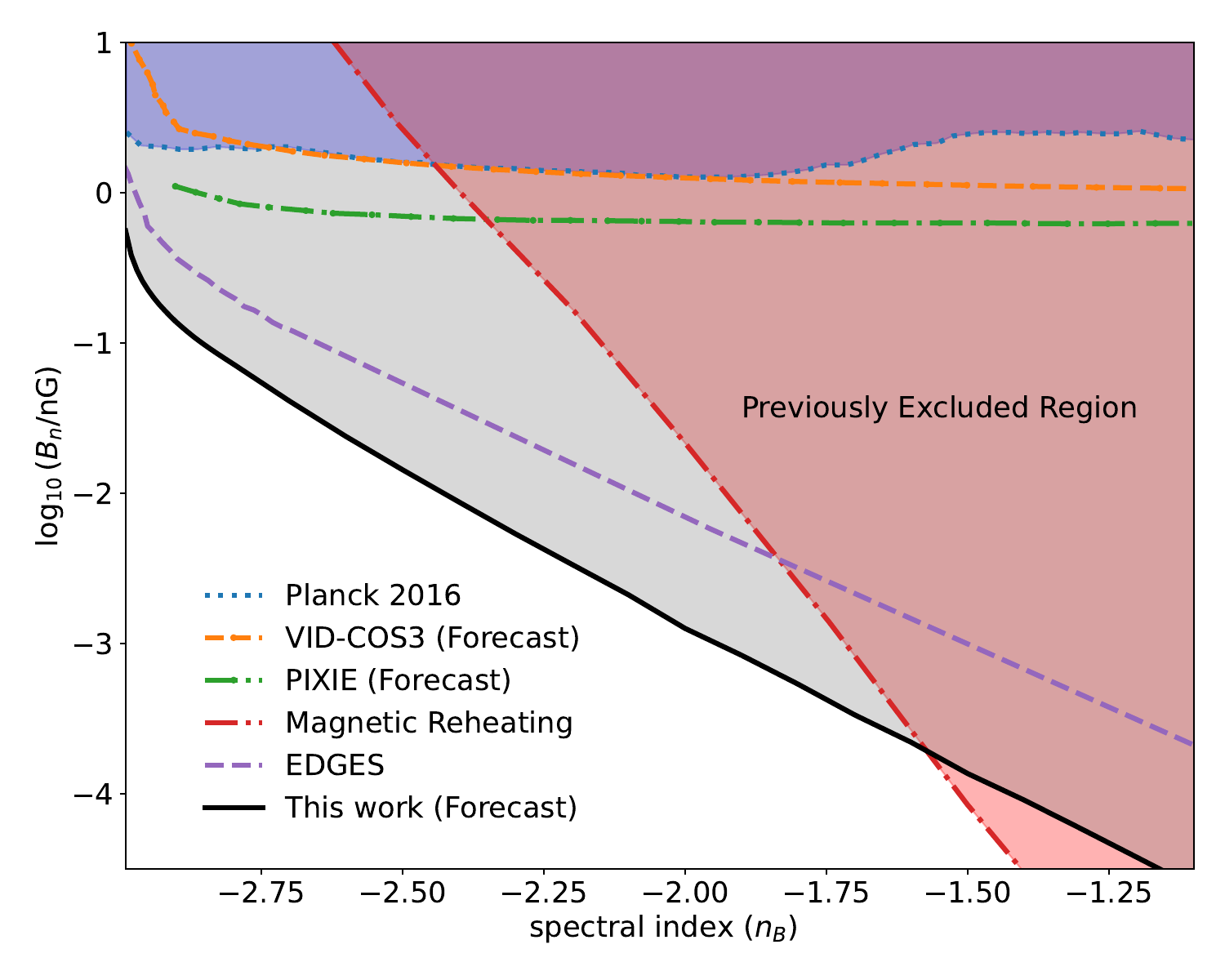}
    \label{fig:figure2b}
    
    \caption{The upper bound on the present-day strength of primordial magnetic fields as a function of spectral index. The red dash-dotted and the blue dotted lines show existing constraints from Planck 2016 and the magnetic reheating process before CMB spectral distortion ($z \gtrsim 2\times10^6$), respectively \cite{refId0, 10.1093/mnrasl/slx195}. In addition, we have also presented some forecasting constraints from future experiments, such as PIXIE (green dot-dashed line) \cite{2014JCAP...01..009K} and VID-COS3 line intensity measurement (orange dot-dashed line) \cite{2023JCAP...09..035A}. The purple dashed line depicts upper bounds on considering $T_g\leq T_{\gamma}$ at $z_{\rm abs} = 17$ in the EDGES framework \cite{2019MNRAS.488.2001M}. The black solid line shows the upper bounds on considering $T_g\leq T_{\gamma}$ at redshifts $z=17$ and $z=89$ together.}
    \label{fig:constrain_plot}
\end{figure}

Lastly, we discuss the existing and forecasted bounds on PMFs. In Fig. (\ref{fig:constrain_plot}), the red dash-dotted line represents the upper bound on PMFs considering ``magnetic reheating'' of plasma before the CMB spectral distortion era \cite{10.1093/mnrasl/slx195}. Here, ``magnetic reheating'' refers to the process where dissipating magnetic energy can alter the baryon-to-photon number ratio. The blue dotted line depicts the bound from Planck 2016 \cite{refId0}. Furthermore, the green dot-dashed and orange dot-dashed lines present the forecasted bounds from PIXIE and VID-COS3, respectively \cite{2014JCAP...01..009K, 2023JCAP...09..035A}. The purple dashed line represents bounds derived by considering $T_g \leq T_{\gamma}$ at $z_{\rm abs} = 17$ within the EDGES framework \cite{2019MNRAS.488.2001M}. The black solid line represents the bound determined from redshift $z = 17$ and $z = 89$ considering $T_{g}\leq T_{\gamma}$. For a nearly scale-invariant magnetic field, the present-day strength should be smaller than $0.55\,\rm nG$. Whereas, for $n_B$ equal to $-2.5$, $-2.0$, $-1.5$, and $-1.0$, the upper limits are $2.2 \times 10^{-2}$, $2.1 \times 10^{-3}$, $2.0 \times 10^{-4}$, and $2.5 \times 10^{-5}\,\rm nG$, respectively. We approximately summarise the bounds as
\begin{equation}
	\log_{10}\left(\frac{B_n}{\text{nG}}\right) \lesssim -\left(2.04\,n_B + 6.8\right)\,.
	\label{constraint}
\end{equation}

\section{Summary and Conclusion}\label{conclusion}

In this work, we studied the effect of primordial magnetic fields on the Dark Ages global 21-cm signal. PMFs can dissipate energy into the intergalactic medium through ambipolar diffusion (Eq. \ref{ambipolar_rate_eqn}) and turbulent decay (Eq. \ref{turb_decay_eqn}) after recombination. Since the Dark Ages signal is independent of astrophysical uncertainties, measuring it can help identify the presence of non-standard physics. Recently proposed experiments, such as FARSIDE \cite{farside}, DAPPER \cite{dapper}, SEAMS \cite{seams}, and LuSee Night \cite{luseenight}, may soon be able to measure the dark ages $T_{21}$ signal.

We then explored the impact of cosmological parameter uncertainties on $T_{21}$ (Eq. \ref{eq: T21}) and $T_{21}^{\prime}$ (Eq. \ref{eq: scaled T21}) from the dark ages. On varying the cosmological parameters $\Omega_mh^2$, $\Omega_bh^2$, and $Y_p$ within their corresponding $68\%$ confidence levels, we found variations of approximately $\sim 6\,\rm mK$ ($-42.5^{+4}_{-2}\,\rm mK$) in global 21-cm signal's amplitude and $\sim 3\,\rm mK$ ($-42.5^{+2}_{-1}\,\rm mK$) in scaled signal's amplitude at $z \simeq 89$ (Fig. \ref{fig:subfig1}). Furthermore, we showed that the shape of the $T_{21}$ signal changes mildly when calculated for different values of cosmological parameters (Fig. \ref{fig:subfig2}). We have also investigated the Dark Ages consistency ratio for PMFs by calculating the corresponding Dark Ages global 21-cm signal at only three different frequencies. This technique could enable future experiments to distinguish between the presence of additional heating sources, such as PMFs, and the standard $\Lambda\rm CDM$ scenario (Fig. \ref{fig:consistency_ratio}). 

We have derived upper bounds on the present-day strength of PMFs with spectral indices $-2.99\leq n_B\leq -1.0$ (Fig. \ref{fig:constrain_plot_a}). Considering $T_g\leq T_{\gamma}$ at $z = 17$ and $z=89$, we obtained maximum values of $B_n$ for different $n_B$ values. We found that for spectral indices in the range from $-1$ to $-2.84$, the 21-cm signal from Dark Ages provides the stronger bounds, while for $-2.85$ to $-2.99$, the signal cosmic dawn provides stronger bounds. We then incorporated X-ray heating using $tanh$ parameterization and re-analysed the bounds. We found comparatively relaxed values of $B_n$ for $n_B\leq -2.85$ in the presence of X-ray heating, while the bounds remained unaffected for $-2.84\leq n_B\leq -1.0$. Therefore, we can conclude that most of the range of the upper bound on $B_n$ is free from astrophysical uncertainties, as they are derived from the Dark Ages era. 

We have presented the obtained bounds compared to the existing ones (Fig. \ref{fig:constrain_plot}). The blue-shaded and red-shaded regions represent the excluded parameter space from Planck 2016 and magnetic reheating, respectively \cite{refId0, 10.1093/mnrasl/slx195}. Whereas the black solid line represents bounds from the dark ages and the cosmic dawn era, without star formation. Compared to magnetic heating, the Dark Ages can provide a stronger and astrophysical uncertainty-free bound on $B_n$ for spectral indices $-2.85\leq n_B\leq -1.58$.

\section*{Acknowledgements}
A. C. N acknowledges financial support from SERB-DST (SRG/2021/002291), India. We thank Rahul Kothari for the fruitful discussion. We thank the anonymous referees for their valuable suggestions and for improving our manuscript.

\bibliography{main}

\end{document}